\documentclass{mn2e}
\usepackage{psfig}
\usepackage{lscape}

\usepackage[totalwidth=510pt, totalheight=700pt]{geometry}

\catcode`\@=11
\def\gsim{\,\mathrel{\mathpalette\@versim>\,}}
\def\lsim{\,\mathrel{\mathpalette\@versim<\,}}
\def\@versim#1#2{\lower 2.9truept \vbox{\baselineskip 0pt \lineskip
    0.5truept \ialign{$\m@th#1\hfil##\hfil$\crcr#2\crcr\sim\crcr}}}
\catcode`\@=12  
\def\Etot{E_{\rm tot}}
\def\Mtot{M_{\rm tot}}

\def\Wstarh{W_{\rm *h}}
\def\av#1{\langle#1\rangle}
\def\mav{\av{m}}
\def\dxcube{d^3{\bf x}}
\def\beq{\begin{equation}}
\def\eeq{\end{equation}}

\def\Nh{N_{\rm h}}

\def\Mstar{M_*}
\def\Nstar{N_*}
\def\Mh{M_{\rm h}}

\def\Phistar{\Phi_*}
\def\betastar{\beta_*}
\def\betah{\beta_{\rm h}}
\def\Phih{\Phi_{\rm h}}
\def\rhostar{\rho_*}
\def\rhomax{\rho_{\rm max}}

\def\rhoh{\rho_{\rm h}}

\def\Mbh{M_{\rm BH}}

\def\Ie{I_{\rm e}}
\def\Re{R_{\rm e}}

\def\Dt{\Delta t}
\def\sae{a_{\rm e}}
\def\sbe{b_{\rm e}}
\def\rh{r_{\rm h}}

\def\rstar{r_{\rm *}}
\def\Rb{R_{\rm b}}

\def\rhalf{r_{\rm M}}
\def\r90{\langle r\rangle _{90}}
\def\r90{\langle r\rangle _{90}}

\def\sg0{\sigma_0}

\def\sglos{\sigma_{\rm los}}
\def\vlos{v_{\rm los}}

\def\tdyn{t_{\rm dyn}}

\def\xvec{{\bf x}}

\def\thetamin{\theta_{\rm min}}

\title[Dissipationless collapse]{Dissipationless collapse, weak
  homology and central cores of elliptical galaxies} 

\author[C. Nipoti, P. Londrillo and L. Ciotti] {Carlo~Nipoti$^1$,
  Pasquale~Londrillo$^2$ and Luca~Ciotti$^{1}$\\ $^1$Dipartimento di
  Astronomia, Universit\`a di Bologna, via Ranzani 1, 40127 Bologna,
  Italy\\ $^2$INAF - Osservatorio Astronomico di Bologna, via Ranzani
  1, 40127 Bologna, Italy}
  
\date{Accepted 2006 May 10.  Received 2006 April 28; in original form
  2005 December 13.}

\pagerange{\pageref{firstpage}--\pageref{lastpage}}
\pubyear{2006}

\begin{document}

\maketitle

\label{firstpage}

\begin{abstract}

By means of high-resolution N-body simulations we revisited the
dissipationless collapse scenario for galaxy formation. We considered
both single-component collapses and collapses of a cold stellar
distribution in a live dark matter halo.  Single-component collapses
lead to stellar systems whose projected profiles are fitted very well
by the Sersic $R^{1/m}$ law with $3.6 \lsim m \lsim 8$.  The stellar
end-products of collapses in a dark matter halo are still well
described by the $R^{1/m}$ law, but with $1.9 \lsim m \lsim 12$, where
the lowest $m$ values are obtained when the halo is dominant.  In all
the explored cases the profiles at small radii deviate from their
global best-fit $R^{1/m}$ model, being significantly flatter. The
break-radius values are comparable with those measured in `core'
elliptical galaxies, and are directly related to the coldness of the
initial conditions. The dissipationless collapse of initially cold
stellar distributions in pre-existing dark matter haloes may thus have
a role in determining the observed weak homology of elliptical
galaxies.

\end{abstract}


\section{Introduction}

It is a well established fact that the end-products of dissipationless
collapses reproduce several structural and dynamical properties of
elliptical galaxies.  For example, the pioneering work of van Albada
(1982, vA82) showed that the end-products of cold collapses have
projected density profiles well described by the $R^{1/4}$ de
Vaucouleurs (1948) law, radially decreasing line-of-sight velocity
dispersion profiles, and radially increasing velocity anisotropy, in
agreement with observations of elliptical galaxies (see also May \&
van Albada 1984; McGlynn 1984). More recently the dissipationless
collapse has been studied in greater detail thanks to the advances in
N-body simulations (e.g., see Aguilar \& Merritt~1990; Londrillo,
Messina \& Stiavelli~1991; Udry~1993; Trenti, Bertin \& van
Albada~2004). These studies show that a smooth final density
distribution with $R^{1/4}$ projected mass profile is produced when
the initial distribution is cold, extended, and clumpy in
phase-space. From the astrophysical point of view the dissipationless
collapse (Eggen, Lynden-Bell \& Sandage 1962) was introduced to
describe a complex physical scenario, in which the gas cooling time of
the forming galaxy is shorter than its dynamical (free-fall) time,
stars form `in flight', and the subsequent dynamical evolution is a
dissipationless collapse.

From the observational point of view, elliptical galaxies are
characterized by a systematic luminosity dependence of their surface
brightness profile, the so-called {\it weak homology}. For example,
high quality photometry of elliptical galaxies indicates that the
$R^{1/m}$ Sersic (1968) law (see equation 9 below), with best-fit
index $m$ ranging from $m \sim 2$ for faint ellipticals up to $m \sim
10$ for the brightest ones, often represents their surface brightness
profiles better than the de Vaucouleurs law (the $m=4$ case; Caon,
Capaccioli \& D'Onofrio 1993; Prugniel \& Simien 1997; Bertin, Ciotti
\& Del Principe 2002; Graham \& Guzm\'an 2003).  In addition, Hubble
Space Telescope observations probing the luminosity profiles of
several ellipticals down to sub-arcsec resolution (Ferrarese et
al.~1994; Lauer et al.~1995) revealed that at these small radii some
profiles flatten (core galaxies), others are characterized by steep
cusps (power-law galaxies).  While the surface brightness profiles of
power-law galaxies are well fitted by the $R^{1/m}$ law down to
centre, within the break radius $\Rb$ the surface brightness profile
of core galaxies stay well below their global best-fit Sersic model
(Graham et al.~2003; Trujillo et al.~2004; Ferrarese et
al.~2006).  In general, core profiles are common among bright
ellipticals, while faint systems tend to have power-law cusps;
remarkably, other galaxy global properties are related to the presence
of the core (e.g. Pellegrini~1999, 2005).

The explanation of the observed weak homology of elliptical galaxies
is important to understand galaxy formation. For example, the presence
of a core is usually interpreted as the signature of the merging of
supermassive black holes, consequence of the merging of the parent
galaxies (e.g. Makino \& Ebisuzaki~1996; Faber et al.~1997;
Milosavljevic \& Merritt~2001), while in N-body simulations of
repeated equal-mass dissipationless (`dry') mergers the best-fit $m$
of the end-products increases with their mass (Nipoti, Londrillo \&
Ciotti 2003a, hereafter NLC03). However, Ciotti \& van Albada~(2001)
and NLC03 showed that repeated dissipationless merging fails to
reproduce the Faber \& Jackson (1976), the Kormendy (1977), and the
$\Mbh$-$\sigma$ (Gebhardt et al.  2000, Ferrarese \& Merritt 2000)
relations, and also that a substantial number of head-on minor
mergings make $m$ decrease, bringing the end-products out of the
edge-on Fundamental Plane.  These results, together with other
astrophysical evidences based on stellar population properties such as
the ${\rm Mg}_2$-$\sigma$ relation (e.g. Burstein et al.~1988; Bender,
Burstein \& Faber~1993), indicate that dry mergings cannot have had a
major role in the formation of elliptical galaxies, and gaseous
dissipation is needed (see, e.g., Robertson et al.~2005; Naab et
al.~2005).

In alternative (or as a complement) to the merging scenario, it is
then of great theoretical interest to explore if (and if yes under
what conditions and to what extent) the dissipationless collapse of
the stellar population produced by a fast episode of gaseous
dissipation and the consequent burst of star formation is able to
reproduce end-products with projected density profiles well described
by the Sersic law. In particular, given that according to the current
cosmological picture galaxies form at peaks of the cold dark matter
distribution (e.g. White \& Rees 1978; White \& Frenk 1991), it is
natural to investigate dissipationless collapse in two-component
systems.  In this work we study this process using high-resolution
two-component N-body simulations, in which the collapse of the stellar
distribution and the response of the dark matter halo are followed in
detail.  As will be shown, dissipationless collapses in pre-existing
dark matter haloes are indeed able to reproduce surprisingly well the
observed weak homology of elliptical galaxies and the flat inner
surface brightness profiles of core ellipticals arise naturally from
dissipationless collapse, with $\Rb$ determined by the coldness of the
initial conditions.

This paper is organized as follows. The numerical simulations are
described in Section~2. The results are presented in
Sections~3. Section~4 summarizes.

\section{Numerical simulations}

\subsection{Initial conditions}
\label{secic}

We consider two classes of simulations. In the first we follow the
virialization of a cold, single-component density distribution. In the
second the initial conditions represent a cold component (stars)
deemed to collapse in a nearly-virialized dark matter halo.

The initial conditions consist of the stellar ($\rhostar$) and the
halo ($\rhoh$) density distributions, and the corresponding virial
ratios $\betastar$ and $\betah$ measure the `coldness' of the
distributions. For example, the stellar virial ratio is defined by
\begin{equation}\label{eqbetastar}
\betastar \equiv {2 K_* \over |U_{*}+\Wstarh|},
\end{equation}
where $K_*$ and
\begin{equation}\label{equstar}
U_*=-\int <\xvec\, , {{\bf \nabla} \Phistar}> \rhostar \dxcube
\end{equation}
are the kinetic energy and the self-gravity of the stellar
distribution,
\begin{equation}\label{eqwstarh}
\Wstarh=-\int <\xvec\, , {{\bf \nabla} \Phih}> \rhostar \dxcube
\end{equation}
is the interaction energy between stars and dark matter, $\Phistar$
and $\Phih$ are the gravitational potentials of stars and dark matter,
respectively, and $<,>$ is the standard inner product.  Note that in
one-component simulations $\betastar=2 K_*/|U_*|$.  All simulations
are evolved up to virialization, which is reached typically within
$40\tdyn$, where
\begin{equation}
\tdyn \equiv {G \Mtot^{5/2} \over (2|\Etot|)^{3/2}}
\end{equation}
is the dynamical time of the system, and $\Mtot$ and $\Etot$ are its
total mass and energy, respectively.

\subsubsection{Stellar component}

The initial configuration of the stellar component is obtained by
introducing inhomogeneity on a smooth, spherically symmetric density
distribution $\rhostar$. In particular, we adopt
\begin{equation}\label{eqplum}
\rhostar(r)={{3 \Mstar  \rstar^2 }\over 4 {\pi (r^2 +\rstar^2)^{5/2}}}
\end{equation}
(Plummer~1911), and
\begin{equation}
\rhostar (r)= {3 - \gamma \over 4\pi}{\Mstar\rstar \over r^\gamma (\rstar+r)^{4-\gamma}}
\label{eqrhosgamma}
\end{equation}
with $\gamma=0$ or $\gamma=1$ (Dehnen~1993; Tremaine et al.~1994),
where $\Mstar$ is the total mass and $\rstar$ the characteristic
radius.  The particles are spatially distributed according to
equations (\ref{eqplum}) or (\ref{eqrhosgamma}) and then randomly
shifted in position (up to $\rstar/5$ in modulus), so that the
distribution results inhomogeneous.

The velocities of the particles are first randomly extracted from an
isotropic Gaussian distribution with a given value of the variance,
and we measure the corresponding virial ratio $\beta_0$. The required
value of the stellar virial ratio $0.002 \lsim \betastar \lsim 0.2$ is
then obtained by scaling the velocity of each particle by
$\sqrt{\betastar/\beta_0}$.  In some case the stellar distribution is
characterized by non-vanishing angular momentum ${\bf J}$, which is
obtained, for fixed $\betastar$, by modifying the direction of the
particles velocity vectors. The system total angular momentum is
quantified by the spin parameter
\begin{equation}
\lambda \equiv {|\Etot|^{3/2} ||{\bf J}|| \over G \Mtot^{5/2}}.
\end{equation}
We consider $\lambda$ in the range\footnote{We recall that typical
values of $\lambda$ inferred from observations are $\sim 10^{-2}$ for
bright ellipticals up to $\sim 10^{-1}$ for faint ellipticals
(e.g. Lake 1983).} $1.7 \times 10^{-2} - 1.3\times 10^{-1}$ (see
Table~1 for details).

\begin{landscape}
\begin{table}
 \centering
  \caption{Simulations parameters.}
  \begin{tabular}{lrrrrrrllrrrrrrrrr}
Name &$\betastar$&$\betah$&$\Mh/\Mstar$&$\rstar/\rh$&$\Nstar$&$\Nh$&$\rhostar$ & $\rhoh$ &$\lambda/10^{-2}$& $c/a$ & $ b/a$ & $m_a$ & $m_b$ & $m_c$& $\epsilon_a$&  $\epsilon_b$&  $\epsilon_c$ \\
\hline
pl04n0002  &  0.002& -  & 0   & -         & 409600 & -       & P & - &0  & 0.47 & 0.97 & $4.3\pm0.06$ &$4.2\pm0.08$  & $3.6\pm0.1$& 0.52&0.54&0.05\\
pl04n0005  &  0.005& -  & 0     & -       & 409600 & -       & P & - &0&0.46 &  0.93  & $4.3\pm0.07$ &$4.2\pm0.1$ &  $3.8\pm0.1$& 0.50& 0.53& 0.09\\
pl1o001  &  0.01& -  & 0   & -          & 102400 & -       & P & -&1.74&0.44 & 0.81 & $4.6\pm0.08$ &$3.7\pm0.2$  & $3.6\pm0.2$& 0.43& 0.59& 0.26\\
pl5o001  &  0.01& -  & 0   & -          & 102400 & -       & P & -&3.63&0.55 & 0.75 & $5.2\pm0.2$ &$4.4\pm0.09$  & $3.7\pm0.2$& 0.29& 0.49& 0.26\\
pl10o001  &  0.01& -  & 0   & -         & 102400 & -       & P & -&3.70&0.56 & 0.75 & $5.0\pm0.2$ &$4.9\pm0.1$  & $3.9\pm0.1$& 0.27& 0.49& 0.28\\
pl20o001  &  0.01& -  & 0   & -         & 102400 & -       & P & -&3.75&0.57 & 0.80 & $4.5\pm0.2$ &$4.4\pm0.1$  & $3.9\pm0.1$& 0.28& 0.47& 0.23\\
pl20f001  &  0.01& -  & 0   & -         & 102400 & -       & P & -&4.98&0.24 & 0.32 & $5.1\pm0.2$ &$4.6\pm0.09$  & $3.9\pm0.09$& 0.38 & 0.53 & 0.23\\
pl04n001  &  0.01 & - & 0   & -         & 409600 & -       & P & -&0&0.47 & 0.97 & $4.2\pm0.07$ &$4.1\pm0.09$  & $3.7\pm0.1$& 0.50& 0.53& 0.05\\
hq2r001   &   0.01& -  & 0   & -         & 204800 & -       & 1 & - & 0 & 0.47 & 0.78 &  $5.7\pm0.3$ &  $5.1\pm0.2$ & $4.7\pm0.2$ & 0.37 & 0.51 &0.23\\
pl20f002  &  0.02& -  & 0   & -         & 102400 & -       & P & -&7.09&0.49 & 0.68 & $4.6\pm0.1$ &$4.6\pm0.1$  & $4.0\pm0.06$& 0.38& 0.53& 0.23\\
pl2r002  &  0.02& -   & 0    & -        & 204800 & -       & P & - &0&0.48 & 0.87  & $4.0\pm0.1$ &$3.8\pm0.1$ & $3.6\pm0.1$ & 0.42& 0.51& 0.13\\
hq2r002  &  0.02& -  & 0     & -       & 204800 & -       & 1 & -&0&0.48 & 0.88  & $6.0\pm0.2$ &$5.7\pm0.3$  & $5.4\pm0.3$& 0.41& 0.49& 0.12\\
g02r002  &  0.02& -  & 0  &  -  &        204800 & -       & 0 & - &0&0.49 & 0.93 & $6.2\pm0.2$ & $6.3\pm0.2$ &  $5.3\pm0.2$& 0.49 & 0.52 & 0.10 \\
pl04n0025  &  0.025& -   & 0    & -        & 409600 & -       & P &-&0&0.46 & 0.63 & $4.5\pm0.3$ &$3.8\pm0.1$  & $3.8\pm0.2$& 0.22& 0.50& 0.40\\
pl20c005  &  0.05& -  & 0   & -         & 102400 & -       & P & -&12.6& 0.49& 0.62 & $4.7\pm0.1$ &$3.8\pm0.1$  & $4.0\pm0.05$& 0.20& 0.53& 0.41\\
pl1n005  &  0.05& -  & 0     & -       & 1024000 & -       & P & - &0&0.40 &  0.66  & $4.4\pm0.2$ &$4.0\pm0.1$ &  $3.7\pm0.2$& 0.35& 0.58& 0.37\\
pl04n005  &  0.05& -  & 0     & -       & 409600 & -       & P & - &0&0.41 &  0.70  & $4.3\pm0.2$ &$4.0\pm0.2$ &  $3.6\pm0.2$& 0.38& 0.57& 0.33\\
hq2r005  &  0.05& -  & 0     & -       & 204800 & -       & 1 & - &0& 0.41 & 0.44 & $8.1\pm0.5$ &$6.2\pm0.2$  & $6.2\pm0.2$& 0.11& 0.62& 0.58\\
pl04n01  &  0.1& -  & 0   & -         & 409600 & -       & P & - &0&0.96 & 0.97 & $4.0\pm0.3$ &$4.2\pm0.3$  & $3.8\pm0.3$& 0.01& 0.04& 0.03\\
hq2r01  &  0.1& -  & 0       & -     & 204800 &  -      & 1 &  - &0&	0.39 & 0.40 & $7.4\pm0.6$ & $5.3\pm0.2$ & $5.2\pm0.2$ & 0.02 & 0.63 & 0.64\\
g02r01  &  0.1& -  & 0       & -     & 204800 &  -      & 0 &  - &0&	0.42 & 0.46 & $7.5\pm0.5$ & $5.0\pm0.1$ & $5.2\pm0.2$ & 0.08 & 0.59 & 0.56 \\
pl04n02  &  0.2& -  & 0   & -         & 409600 & -       & P & - &0&0.96 & 0.97 & $3.6\pm0.3$ &$3.7\pm0.3$  & $4.0\pm0.4$& 0.01& 0.01& 0.01\\
hq2r02  &  0.2 & - & 0      &  -    & 204800 & -       & 1 & - &0& 0.98& 0.99  &	$5.6\pm0.3$& $5.6\pm0.3$ & $5.6\pm0.3$ & 0.01 & 0.02 & 0.01\\
hpl4r05m001  &  0.01& 0.7  & 0.5 & 4 & 163840 & 81920  & P & 1 &0&  0.28  & 0.40 & $4.3\pm0.2$ &$3.1\pm0.1$  & $3.1\pm0.1$& 0.35& 0.75& 0.63\\
hpl4r1m001   &  0.01& 0.8  & 1   & 4 & 102400 & 102400  & P & 1 &0& 0.24  & 0.34  & $3.7\pm0.1$ &$2.8\pm0.07$  &  $2.8\pm0.06$& 0.31 & 0.77 & 0.68 \\
hpl20f4r1m001&  0.01& 0.8  & 1   & 4 & 102400 & 102400  & P & 1&3.14&0.24 & 0.32 & $3.8\pm0.1$ &$2.7\pm0.07$  & $2.7\pm0.06$& 0.28 & 0.79 & 0.71 \\
g0pl04r1m001 &  0.01& 0.3  & 1   & 0.4  & 102400 & 102400  & P & 0 &0& 0.37 & 0.57 & $3.3\pm0.05$ &$2.9\pm0.06$  & $2.9\pm0.05$& 0.35& 0.64& 0.46\\
g0pl08r1m001 &  0.01& 0.3  & 1   & 0.8  & 102400 & 102400  & P & 0 &0& 0.43 & 0.87 & $4.3\pm0.09$ &$3.7\pm0.08$  & $3.6\pm0.1$& 0.48& 0.56& 0.13\\
hpl025r2m002 &  0.02& 0.4  & 2   &0.25&102400 & 204800  & P & 1 &0& 0.34 & 0.35& $3.4\pm0.1$ &$3.1\pm0.03$ &   $3.0\pm0.03$  & 0.0 & 0.68& 0.68\\
plpl05r2m002 &  0.02& 0.1  & 2   & 0.5 & 204800 & 409600   & P & P &0& 0.30 & 0.34 & $4.0\pm0.1$ &$3.4\pm0.04$  &  $3.9\pm0.1$& 0.22& 0.71& 0.71\\
plpl05r5m002 &  0.02& 0.02 & 5   & 0.5 & 81920 & 409600   & P & P &0& 0.47 & 0.72  & $4.8\pm0.1$ &$4.0\pm0.1$ & $4.8\pm0.1$ & 0.32& 0.49& 0.29\\
g0hq05r2m002 &  0.02& 1    & 2   & 0.5&204800 & 204800  & 1 & 0 & 0 &0.42 &0.85 &  $5.4\pm0.2$ & $5.6\pm0.3$ & $5.2\pm0.2$ & 0.49 & 0.57 & 0.17\\
hus20r1m002  &  0.02& 0.9  & 1   & 20 & 102400 & 102400  & U & 1 &0&  0.87 & 0.90 & $5.6\pm0.4$ &$6.5\pm0.6$  & $5.6\pm0.4$& 0.01& 0.11& 0.11\\
hpl4r1m005   &  0.05& 0.8  & 1   & 4 & 102400 & 102400  & P & 1 &0& 0.98 & 0.99  & $2.5\pm0.03$ &$2.5\pm0.04$  &  $2.5\pm0.04$& 0.01& 0.01& 0.01\\
hpl4r2m005   &  0.05& 0.8  & 2   & 4 & 102400 & 204800 & P & 1 &0& 0.99 & 1.00  & $2.2\pm0.03$ &$2.3\pm0.04$  &$2.2\pm0.03$& 0.01& 0.00& 0.00\\
hpl8r2m005   &  0.05& 0.9  & 2   & 8 & 102400 & 204800  & P & 1 &0& 0.99 & 1.00  & $2.0\pm0.03$ &$1.9\pm0.02$  &$1.9\pm0.02$& 0.00& 0.01& 0.01\\
hpl2r4m005   &  0.05& 0.8  & 4   & 2 & 81920 & 327680   & P & 1 &0&0.99 & 1.00 & $2.1\pm0.03$ &$2.1\pm0.03$  & $2.1\pm0.03$& 0.00& 0.01& 0.00\\
npl4r2m005   &  0.05& 0.9  & 2   & 4 & 102400 & 204800  & P & N &0& 0.33 & 0.33 & $3.4\pm0.2$ &$2.8\pm0.09$  &  $2.8\pm0.08$& 0.01& 0.01& 0.01\\
npl4r4m005   &  0.05& 1    & 4   & 4 & 81920 & 327680   & P & N &0& 0.97 & 0.98 & $2.4\pm0.05$ &$2.3\pm0.04$  &  $2.3\pm0.05$& 0.01& 0.00& 0.00\\
g15pl4r2m005 &  0.05& 1   & 2   & 4 & 102400 & 204800   & P & 1.5 &0& 0.98 & 0.99 & $1.9\pm0.04$ &$2.0\pm0.05$  & $2.0\pm0.04$& 0.01& 0.02& 0.00\\
g0pl4r1m005  &  0.05& 0.7 &  1  & 4 & 102400 & 102400   & P & 0 &0& 0.33 & 0.33 & $3.2\pm0.2$ &$2.8\pm0.06$  & $2.7\pm0.07$& 0.01& 0.70& 0.70\\
g0hq05r1m005 &  0.05& 0.2 &  1  & 0.5 & 102400 &  102400 & 1 & 0 &0& 0.39 & 0.39 & $12.1\pm1.2$ &$7.3\pm0.3$ & $7.8\pm 0.4$&  0.03 & 0.64 & 0.63\\
hpl2r2m01    &  0.1& 0.8  &  2  & 2 & 102400 & 204800  & P & 1 &0&0.98 & 0.99 &$2.6\pm0.1$ &$2.6\pm0.1$  &$2.6\pm0.1$& 0.01& 0.03& 0.02\\
\hline
\end{tabular}

\medskip

\flushleft{First column: name of the simulation. $\betastar$ and
  $\betah$: initial stellar and halo virial ratios,
  respectively. $\Mh/\Mstar$: halo to stellar mass
  ratio. $\rstar/\rh$; ratio of initial characteristic radii. $\Nstar$
  and $\Nh$: number of stellar and halo particles.  $\rhostar$ and
  $\rhoh$: initial stellar and halo distribution (P, N, U: Plummer,
  NFW, and uniform spheres distributions, respectively.  Numbers
  identify $\gamma=0$, 1, 1.5 models.). $\lambda$: spin parameter.
  $c/a$ and $b/a$: minor-to-major and intermediate-to-major axis
  ratios. $m_a$, $m_b$, $m_c$ and $\epsilon_a$, $\epsilon_b$,
  $\epsilon_c$: best-fit Sersic indices and ellipticities for
  projections along the principal axes.}

\end{table}
\end{landscape}

\subsubsection{Halo component}

In most of the simulations the dark matter halo component is
represented by a $\gamma$-model distribution 
\begin{equation}
\rhoh (r)= {3 - \gamma \over 4\pi}{\Mh\rh \over r^\gamma (\rh+r)^{4-\gamma}},
\label{eqrhoh}
\end{equation}
where $\Mh$ is the total mass, $\rh$ is the characteristic radius, and
we adopt $\gamma=0$, $1$, or $1.5$.  In two simulations the halo
component is represented by a Navarro, Frenk \& White (1996, NFW)
density profile with scale radius $\rh$, truncation radius $10\rh$,
and total mass $\Mh$.  Note that for the $\gamma=1$ model
(Hernquist~1990) $\rhoh (r) \sim r^{-1}$ for $r \to 0$, as in the NFW
density profile.  The velocity distribution of the halo particles is
isotropic and, with few exceptions, such that the halo would be in
equilibrium in the absence of the stellar component. In practice, in
most cases the halo is nearly-virialized ($0.7 \lsim \betah \lsim 1$),
while in a subset of simulations the halo is strongly out of
equilibrium ($\betah \ll 1$; see Table~1 and Section~3.4). Length-scale
ratios $\rstar/\rh$, and mass ratios $\Mh/\Mstar$ characterizing the
initial conditions are listed in Table~1.

\begin{figure}
\psfig{file=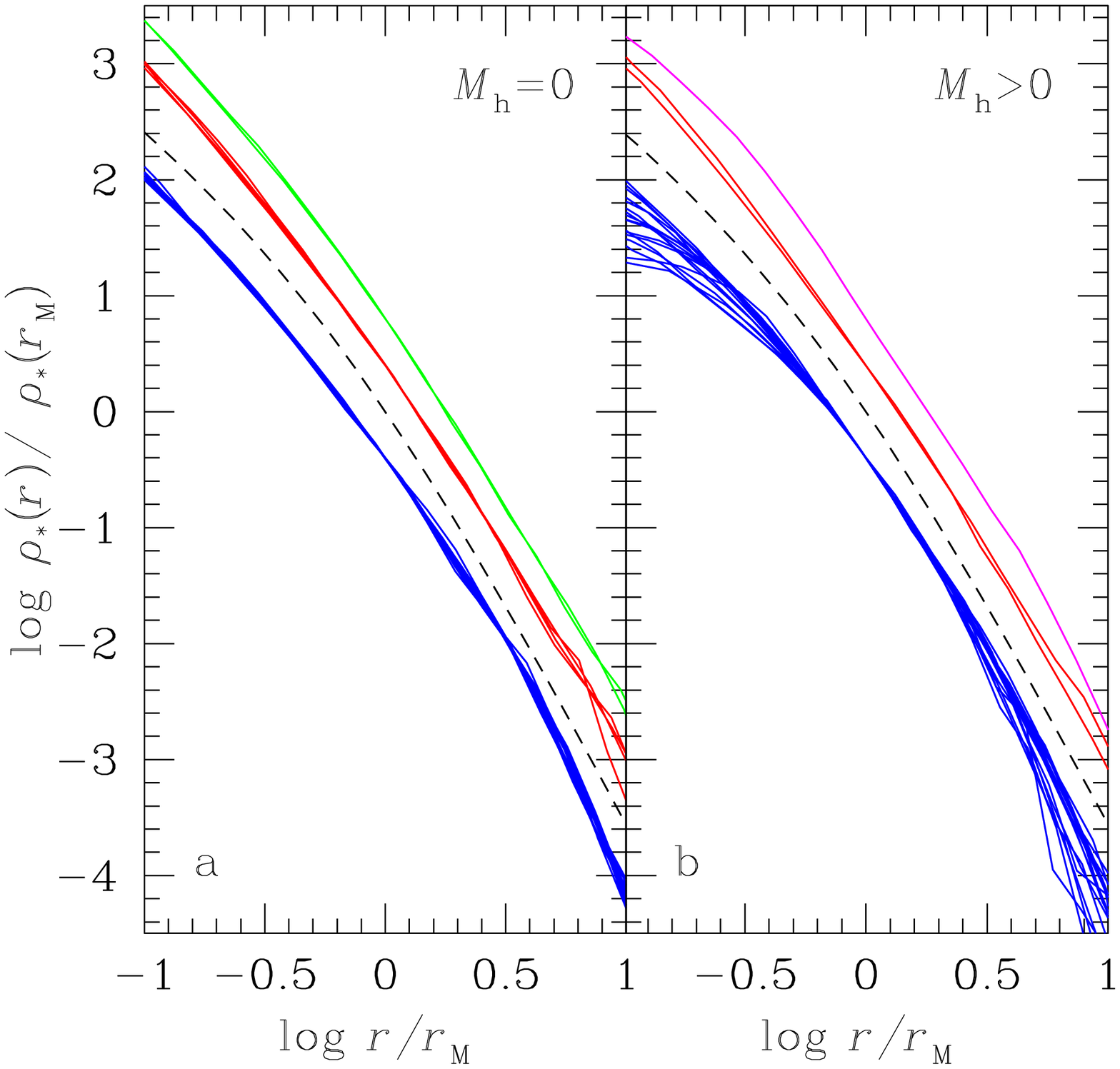,width=.9\hsize}
\psfig{file=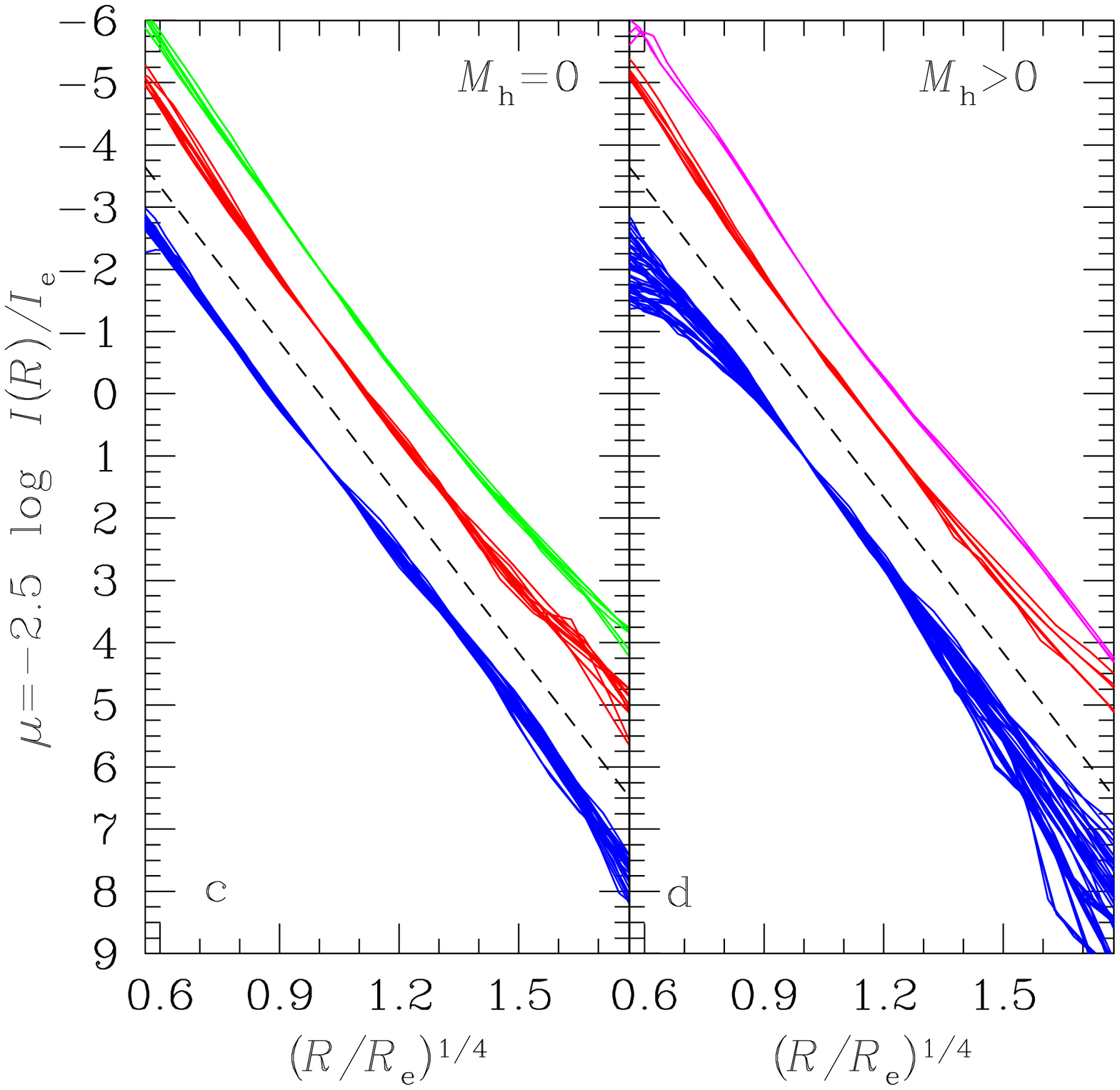,width=.9\hsize}
 \caption{Angle-averaged spatial density profile of the end-products
   of one-component collapses (panel a; solid lines) and of the
   stellar component of two-component collapses (panel b; solid
   lines).  The dashed line is the $\gamma=1.5$ profile. Panels c and
   d: projected density profiles (along the principal axes) of the
   stellar component of the end-products of one- and two-component
   collapses. The dashed line represents the $R^{1/4}$ law. Different
   colours indicate different initial stellar density profiles:
   Plummer (blue), $\gamma=1$ model (red), $\gamma=0$ model (green),
   hus20r1m002 (magenta, see Table~1). Blue, red, green and magenta
   curves are vertically offset by $-0.3$, $0.3$, $0.6$ and $0.6$ dex
   (top panels), and by $1$, $-1$, $-2$, and $-2$ magnitudes (bottom
   panels), respectively.}
\label{figden}
\end{figure}

\subsection{Numerical methods}
\label{secnum}

Numerically, the initial distribution of particles in phase-space is
realized as described in Nipoti, Londrillo \& Ciotti (2002).  For the
stellar distribution we use typically $\Nstar \sim 1-4 \times 10^5$
particles, and in the two-component cases the number of
particles $\Nh$ of the dark matter distribution is such that halo and
stellar particles have the same mass (with few exceptions in which
halo particles are twice as massive as stellar particles; see
Table~1).

For the simulations we used the parallel N-body code FVFPS (Fortran
Version of a Fast Poisson Solver; Londrillo, Nipoti \& Ciotti 2003;
NLC03), based on Dehnen's (2000, 2002) scheme. We adopt the following
values for the code parameters: minimum value of the opening parameter
$\thetamin=0.5$, and softening parameter $\varepsilon=0.005-0.025$ in
units of $\rstar$ (depending on the number of particles). The
time-step $\Dt$, which is the same for all particles, is allowed to
vary adaptively in time as a function of the maximum particle density
$\rhomax$. In particular, we adopted $\Dt=0.3/\sqrt{(4\pi G
\rhomax)}$, the classical time-step criterion for the stability of the
leap-frog time integration. This gives time-steps as small as $6
\times 10^{-4}\tdyn$ during the first phases of the coldest collapses,
assuring energy conservation to within $10^{-3}-10^{-4}$.  In order
to estimate discreteness effects, we ran several test simulations
varying the softening length $\varepsilon$ and the number of
particles. We found that the main properties of the end-product of a
simulation with $\Nstar \sim 10^6$ are not significantly different
from those of lower-resolution simulations with the same initial
conditions. In particular, the tests indicate that, for a
one-component simulation with $\Nstar \sim 4 \times 10^5$ and
$\varepsilon=0.01\rstar$, the final density distribution can be
trusted down to radii $\sim 2 \varepsilon$ ($\sim\,\,0.01\,\rhalf$,
where $\rhalf$ is the half-mass radius), in accordance with
numerical convergence studies (e.g. Power et al.~2003).

The intrinsic and projected properties of the end-products are
determined following Nipoti et al.~(2002), while the position of the
centre of the system is determined using the iterative technique
described by Power et al.~(2003).  In particular, we measure the axis
ratios $c/a$ and $b/a$ of the inertia ellipsoid (where $a$, $b$ and
$c$ are the major, intermediate and minor axis) of the stellar
component, their angle-averaged density distribution and half-mass
radius $\rhalf$. For each end-product, in order to estimate the
importance of projection effects, we consider three orthogonal
projections along the principal axes of the inertia tensor, measuring
the ellipticity $\epsilon=1-\sbe/\sae$, the circularized projected
density profile and the circularized effective radius
{$\Re\equiv\sqrt{\sae\sbe}$} (where {$\sae$} and {$\sbe$} are the
major and minor semi-axis of the effective isodensity ellipse).  We
fit (over the radial range $0.1 \, \lsim \, R/\Re \, \lsim \, 10$) the
circularized projected stellar density profile of the end-products
with the $R^{1/m}$ Sersic~(1968) law:
\begin{equation}\label{eqser}
I(R)=\Ie \,
\exp\left\{-b(m)\left[ \left( \frac{R}{\Re} \right)^{1/m} -1 \right]\right\},
\end{equation}
where $\Ie\equiv I(\Re)$ and $b(m)\simeq 2m-1/3+4/(405m)$ (Ciotti \&
Bertin 1999).  In the fitting procedure $m$ is the only free
parameter, because $\Re$ and $\Ie$ are determined by their {\it
measured} values obtained by particle count. 

\begin{figure}
\psfig{file=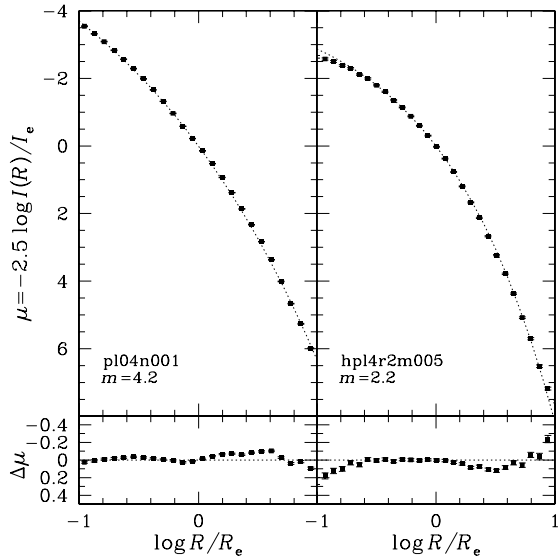,width=.9\hsize}
 \caption{Circularized projected stellar density profiles of the
 end-products of representative one-component (left) and
 two-component (right) collapses.  The dotted lines are the
 best-fitting Sersic models. 1-$\sigma$ error bars are also
 plotted. }\label{figdeltamu}
\end{figure}

\section{Properties of the end-products}

\subsection{Density distribution}

\subsubsection{Spatial density profile}

In Fig.~\ref{figden}a we plot the angle-averaged density profiles of
the end-products of one-component collapses (solid lines). The
relatively small spread of the curves in the diagram indicates that
the profiles are similar over most of the radial range ($0.1 \lsim
r/\rhalf \lsim 10$), independently of the details of the initial
density distribution and of the specific angular momentum. In
particular the profiles are reasonably well described by the
$\gamma=1.5$ model (dashed line). 

In Fig.~\ref{figden}b we plot the angle-averaged density profiles of
the stellar component of two-component simulations. The final stellar
density profiles of two-component collapses are more varied than the
results of one-component collapses, and they are typically flatter at
$r<\rhalf$ and steeper at $r>\rhalf$ than the $\gamma=1.5$ model.  So
the overall (and expected) effect of the presence of the halo in the
collapse is to introduce significant structural homology.

The initial and final halo density distributions are very similar when
the halo is dominant ($\Mh/\Mstar\ge2$). When $\Mh/\Mstar\le1$ the
halo profile in the central regions is significantly modified by the
collapse. In particular, the final inner halo profile is shallower
than the initial one if $\rstar/\rh \gsim 1$ at $t=0$, and steeper if
$\rstar/\rh \lsim 1$.  Thus, the collapsing stellar component may be
able to modify the density distribution of a pre-existing
halo\footnote{Analytical restrictions on the relative density radial
trend of stars and dark matter have been derived in Ciotti \&
Pellegrini~(1992), and Ciotti~(1996; 1999).}. The final total (stellar
plus halo) density profiles are characterized by significant
structural non-homology, and are typically well represented by
$\gamma$-models with $0\lsim \gamma \lsim 2$.

Finally, we note that in both one- and two-component simulations
the final density distribution tends to be steeper when the initial
stellar distribution is a $\gamma=0$ (green curves in
Fig.~\ref{figden}) or a $\gamma=1$ (red curves) model than for Plummer
initial conditions (blue curves; see also Section~3.4).

\begin{figure}
\psfig{file=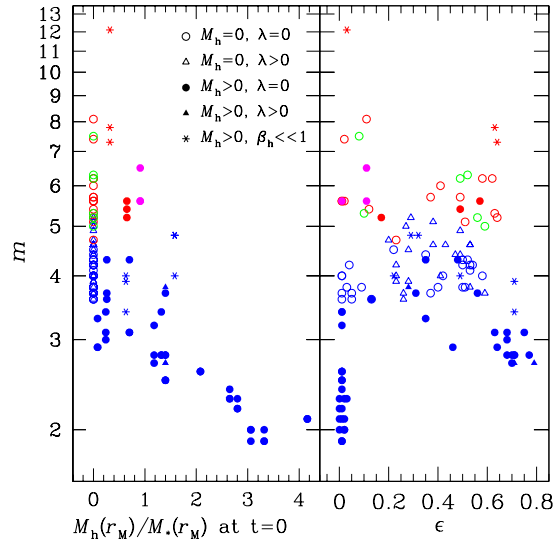,width=.9\hsize}
 \caption{Left: best-fit Sersic parameter $m$ as function of the
 initial dark-to-luminous mass ratio $\Mh(\rhalf)/\Mstar(\rhalf)$
 within the half-mass radius of the stellar distribution. Right:
 best-fit Sersic index as a function of the ellipticity. For each
 model we plot three points corresponding to the three principal
 projections. Colours have the same meaning as in Fig.~\ref{figden}.}
\label{figmem}
\end{figure}

\subsubsection{Projected density profile: Sersic fit}

The projected density profiles of the stellar end-products are
analysed as described in Section~\ref{secnum}.  The fitting radial
range $0.1\,\lsim\, R/\Re\,\lsim\, 10$ is comparable with or larger
than the typical ranges spanned by observations (e.g.
Bertin~et~al.~2002). The best-fit Sersic indices $m_a$, $m_b$ and
$m_c$ (for projections along the axes $a$, $b$, and $c$, respectively)
are reported in Table~1, together with the $1\sigma$ uncertainties
corresponding to $\Delta \chi^2=1$. We note that the relative
uncertainties on the best-fit Sersic indices are in all cases smaller
than 10 per cent.

Figure~\ref{figden}c shows the circularized projected density profiles
of the end-products of one-component simulations (three projections
for each system, each normalized to the corresponding value of $\Ie$),
while the de Vaucouleurs profile is the dashed straight
line. Apparently all the end-products of one-component collapses do
not deviate strongly from the $R^{1/4}$ law over most of the radial
range, in agreement with previous studies (vA82, Londrillo et
al. 1991, Trenti et al. 2004). In particular, the Sersic index is
found in the range $3.6 \lsim m \lsim 8$.  The quality of the fits is
apparent in Fig.~\ref{figdeltamu} (left), which plots the surface
brightness profile of a projection of one of these end-products,
together with the best-fit ($m=4.2\pm0.07$) Sersic law, and the
corresponding residuals. The average residuals between the data and
the fits are typically $ 0.04\lsim\langle\Delta\mu\rangle \lsim 0.2$,
where $\mu=-2.5 \log I(R)/\Ie$.

\begin{figure}
\psfig{file=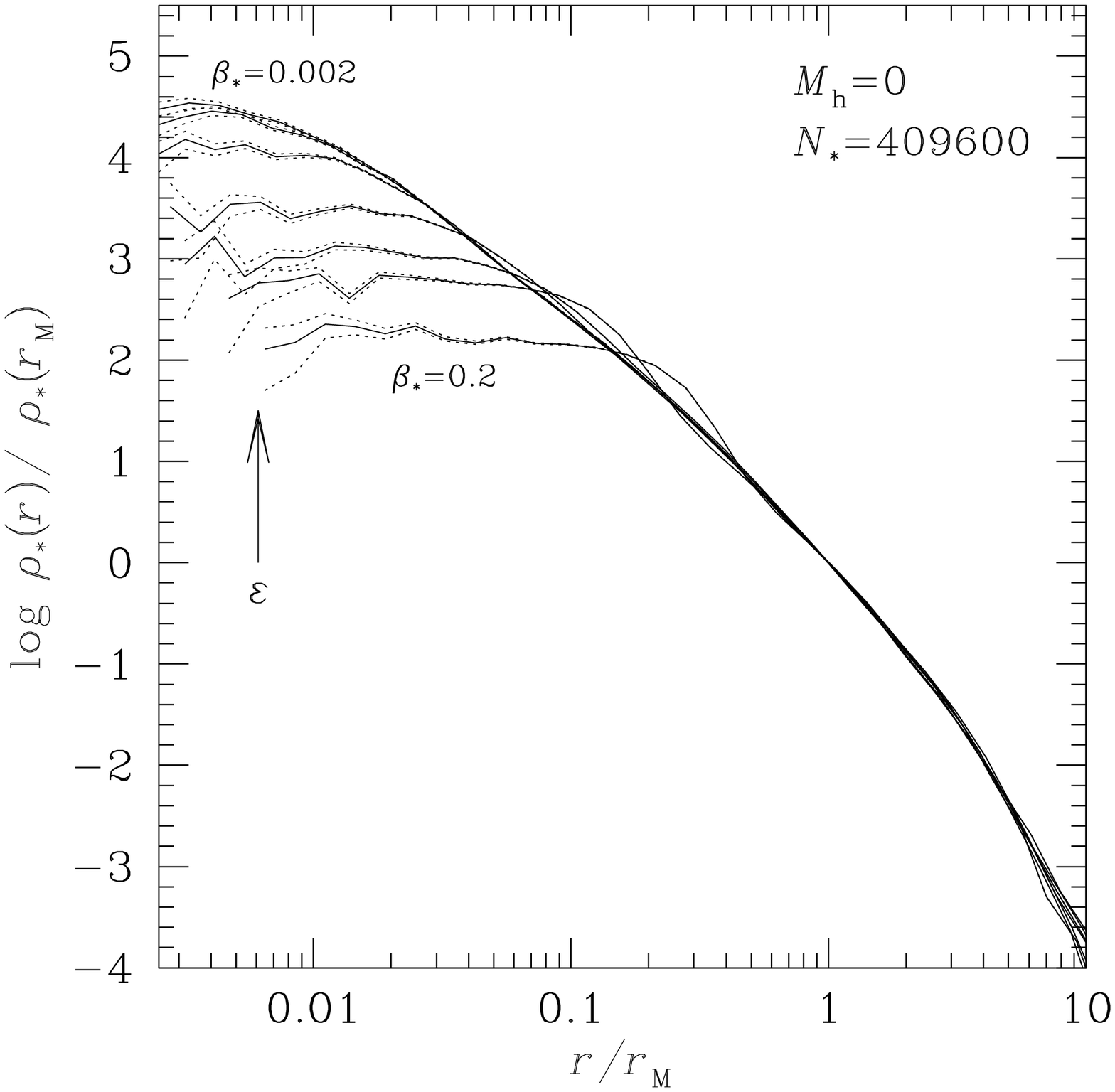,width=.9\hsize}
\psfig{file=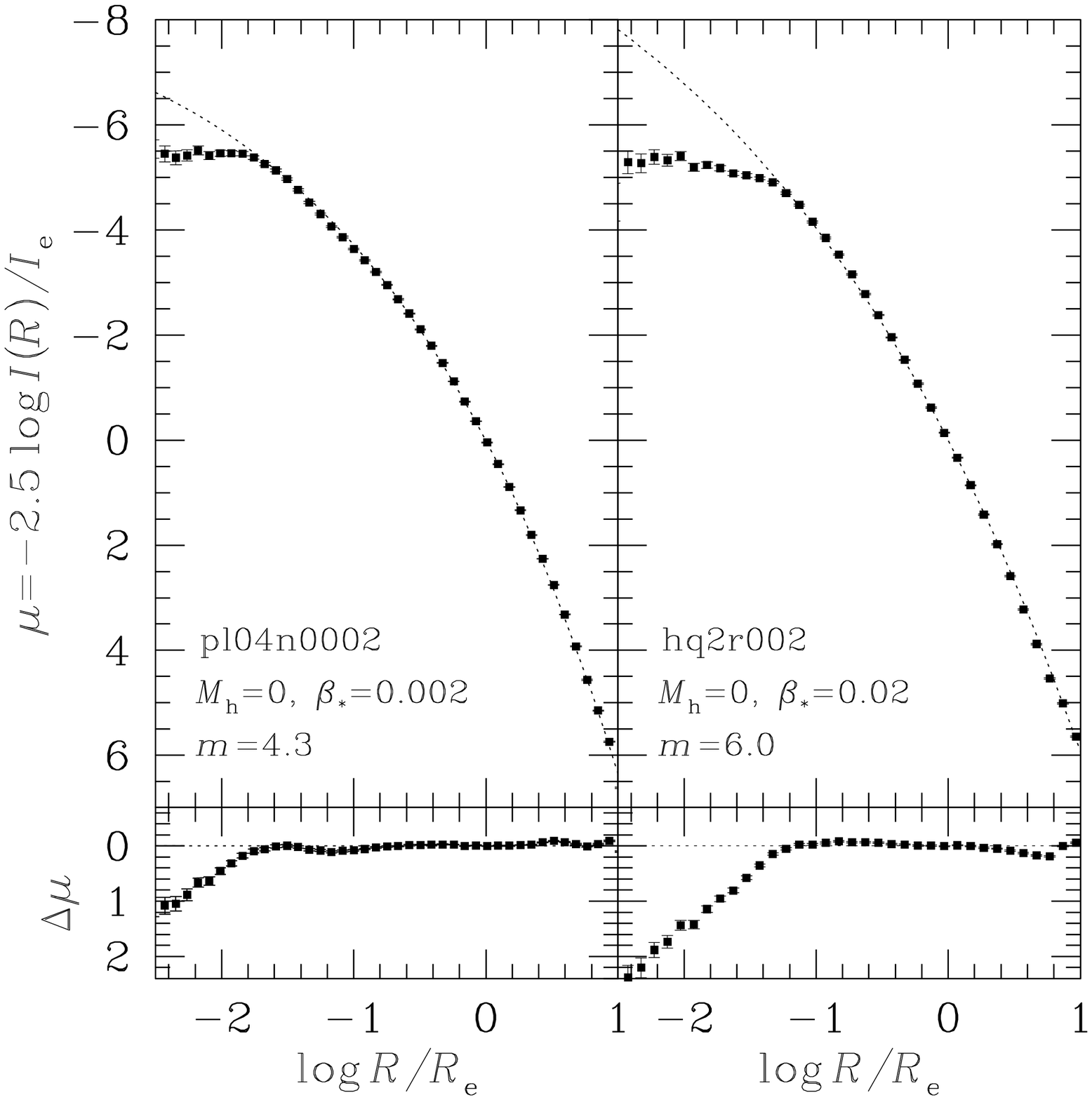,width=.9\hsize}
 \caption{Top: angle-averaged density profiles of the end-products of
   one-component simulations with $\betastar=0.002$, 0.005, 0.01,
   0.025, 0.05, 0.1 and 0.2 (solid lines).  The arrow indicates the
   typical simulation softening-length $\varepsilon$, while dotted
   curves indicate 1-$\sigma$ uncertainties due to statistical noise.
   Bottom: circularized projected density profiles of two
   representative one-component collapses. The bars are 1-$\sigma$
   uncertainties. The dotted lines are the best-fitting Sersic models
   over the radial range $0.1<R/\Re<10$.  }
\label{figinn}
\end{figure}

In Fig.~\ref{figden}d we finally plot the circularized projected
stellar density profiles of the end-products of two-component
simulations.  These systems deviate systematically from the $R^{1/4}$
(straight line), and in most cases the profile remain below it at
small and large radii. For these systems we found $1.9\lsim m
\lsim12$, with average residuals in the same range as those of
one-component collapses. We note that values of $m$ significantly
larger than 4 are found only for collapses starting from sufficiently
concentrated stellar distributions (see
Section~3.4). Figure~\ref{figdeltamu} (right) plots the projected
profile of a representative two-component simulation with
$\Mh/\Mstar=2$ together with the best-fit $(m=2.2\pm0.03)$ model and
the residuals.

The situation is summarized in the left panel of Fig.~\ref{figmem},
where it is apparent how the initial amount and concentration of the
dark matter is crucial in determining the final shape of the profile.
In particular, the trend is that $m$ is anti-correlated with the
initial ratio $\Mh(\rhalf)/\Mstar(\rhalf)$ within $\rhalf$.  However,
it is also apparent from Fig.~\ref{figmem} that there is significant
spread in $m$ for similar values of $\Mh(\rhalf)/\Mstar(\rhalf)$,
which reflects differences in the initial stellar density profiles and
halo virial ratios (see Section~3.4). We also note that for some
strongly aspherical end-products $m_a$, $m_b$, and $m_c$ are
significantly different (see Table~1).

\begin{figure}
\psfig{file=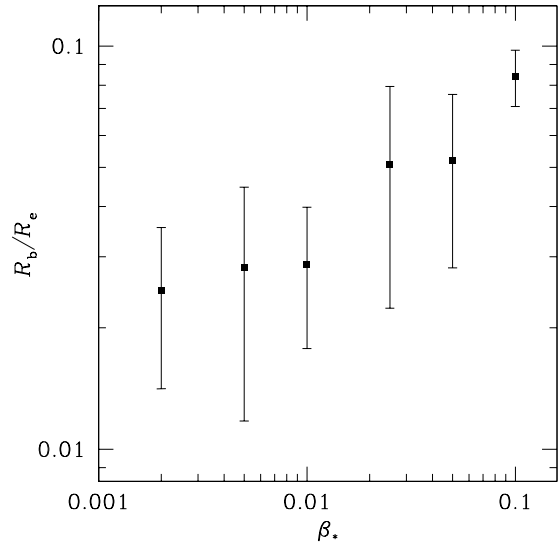,width=0.9\hsize}
 \caption{Break radius normalized to the effective radius as a
 function of the initial virial ratio for the simulations in
 Fig.~\ref{figinn} (top) with $\betastar\le0.1$. The symbols refer to
 the average among the three considered projections values, which span
 the range represented by the vertical bars.}
\label{figbetarb}
\end{figure}

\begin{figure}
\psfig{figure=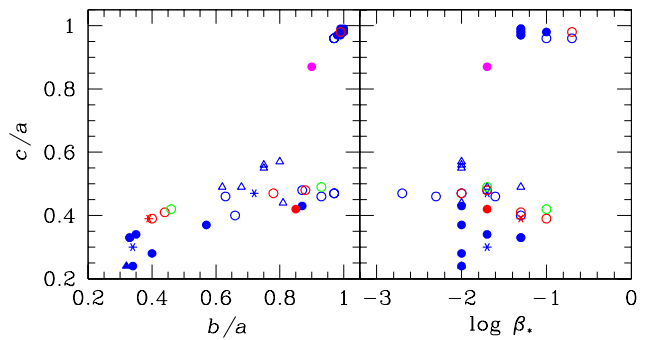,width=\hsize,angle=0,bbllx=18bp,bblly=143bp,bburx=577bp,bbury=442bp,clip=}
 \caption{Final axis ratios $c/a$ as a function of $b/a$ (left), and
 of the initial stellar virial ratio $\betastar$ (right). The meaning
 of symbols and colours is the same as in Fig.~\ref{figmem}.}
\label{figaxes}
\end{figure}

\begin{figure*}
\centerline{\psfig{file=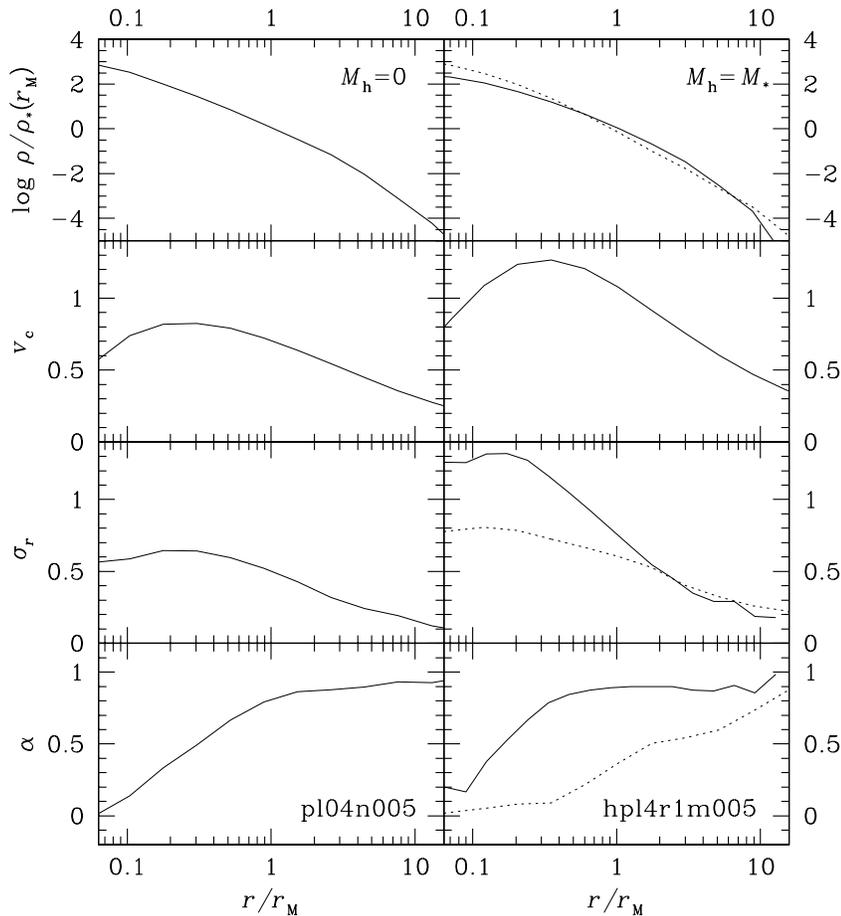,width=0.7\hsize}}
 \caption{Density, circular velocity, radial velocity dispersion, and
 anisotropy parameter for the end-products of a one-component
 simulation (left panels) and a two-component simulation (right
 panels). Solid lines refer to stars; dashed lines to dark matter.
 Velocities are normalized to $\sqrt{G\Mstar/\rhalf}$. }
\label{fig3d}
\end{figure*}

\subsubsection{Behaviour at small radii}
\label{secbeh}

So far we have considered the properties of intrinsic and projected
density profiles at radii $\gsim 0.1\rhalf$ (and 0.1$\Re$).  We now
focus on the behaviour of the profiles at smaller radii, yet confining
our discussion to $r \gsim\,0.01\,\rhalf$ where artificial numerical
effects do not affect the profile (see Section~\ref{secnum}).

As apparent from Fig.~\ref{figinn} (top), the end-product density
profiles of one-component collapses have flat cores at $r \lsim 0.1
\rhalf$, in agreement with previous studies (vA82; May \& van Albada
1984).  Correspondingly, the projected density profiles are
characterized by a break radius $\Rb$, in the sense that for $R<\Rb$
they stay below the best-fit Sersic profiles over $0.1<R/\Re<10$, as
shown in Fig.~\ref{figinn} (bottom panels), for a case with
$\betastar=0.002$ (left) and a case with $\betastar=0.02$ (right).
Thus, $\betastar$ determines the radial range over which the final
surface density profile is well fitted by the Sersic law, with colder
initial conditions producing smaller cores (see Fig.~\ref{figbetarb}).
Note that we do not represent $\Rb/\Re$ for $\betastar=0.2$, because in
this case the core extends to $R\gsim0.1\Re$, affecting the Sersic fit
in the considered radial range $0.1<R/\Re<10$, so the break radius is
not well defined.

Also the projected density profiles of two-component end-products
flatten at small radii and deviate from an inwards extrapolation of
the best-fitting Sersic law.  However, the flattening is typically
more gradual and the break is not as apparent as in the one-component
cases. This is in part due to the fact that the stellar end-products
of two-component simulations are characterized in general by smaller
$m$, corresponding to quite shallow profiles.

\subsubsection{Intrinsic shape and projected ellipticity}

Figure~\ref{figaxes} shows the location of the end-products in the
($b/a$, $c/a$) and ($\betastar$, $c/a$) planes.  The final states of
one-component collapses are roughly spherical for $\betastar \gsim
0.1$, prolate ($c/a \sim b/a \sim 0.3$) for $0.05 \lsim \betastar
\lsim 0.1$ and oblate ($c/a \sim c/b \sim 0.5$) for $\betastar \lsim
0.05$, in agreement with the findings of previous works (Aguilar \&
Merritt~1990; Londrillo et al.~1991).  A similar behaviour is found in
two-component simulations, though in this case the halo mass and
concentration, beside $\betastar$, play a role in determining the final
shape of the stellar component. In particular, in case of dominant
halo the remnant is roughly spherical even for $\betastar=0.02-0.05$.
When the stellar component develops a bar, the dark matter halo
becomes mildly flattened ($c/a \sim b/a \sim 0.7-0.8$).

The ellipticity $\epsilon$ (measured for each of the principal
projections of each end-product) is found in the range $0-0.8$,
consistent with those observed in real ellipticals, and no significant
correlation with the Sersic index $m$ is found (Fig.~\ref{figmem},
right panel).  In particular, $m\gsim 4$ systems span a wide range in
ellipticity, while lower-$m$ systems are found both round and very
flattened (though $m$ as low as 2 is found only for $\epsilon \sim
0$).

\subsection{Kinematics}

The internal dynamics of the end-products is quantified by a `bona
fide' circular velocity $v_c^2 \equiv G \Mtot(r)/r$, the
(angle-averaged) radial component $\sigma_r$ of the velocity
dispersion tensor, and the anisotropy parameter $\alpha(r) \equiv 1
-0.5\sigma^2_{\rm t}/\sigma^2_r$, where $\sigma_{\rm t}$ is the
tangential component of the velocity-dispersion tensor.  These
quantities, together with the corresponding angle-averaged density
distributions, are shown in Fig.~\ref{fig3d} for two representative
simulations.  The radial dependence of $v_c$, $\sigma_r$ and $\alpha$
is qualitatively similar in the one- and two-component cases. The
circular velocity is roughly constant for $0.2 \lsim r/\rhalf \lsim 1$
and decreases inwards and outwards; $\sigma_r$ has a plateau at small
radii and decreases at larger radii. The systems are approximately
isotropic in the centre ($\alpha \sim 0$) and strongly radially
anisotropic for $r\gsim\rhalf$, in agreement with previous results
(e.g. vA82; Trenti et al.~2004).

\begin{figure}
\psfig{file=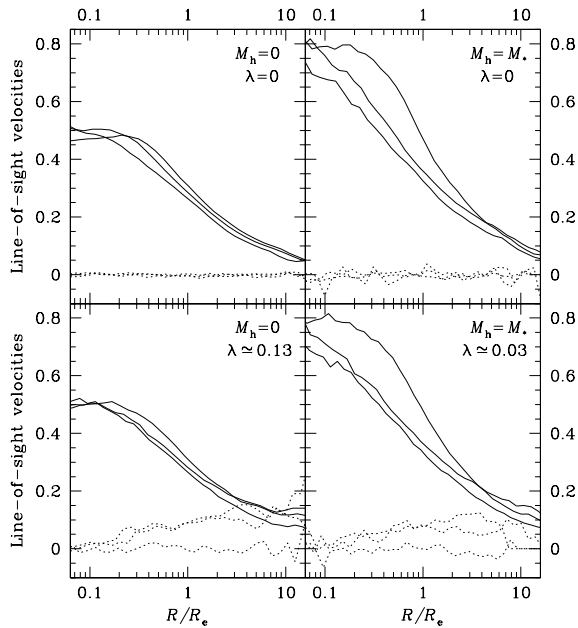,width=\hsize}
\caption{Line-of-sight velocity (dotted line) and velocity dispersion
(solid line), along the principal axes, normalized to
$\sqrt{G\Mstar/\Re}$, for the stellar component of four end-products
with different values of $\Mh$ and $\lambda$.}
\label{fig2d}
\end{figure}

For each model projection we compute the line-of-sight streaming
velocity $\vlos$ and the line-of-sight velocity dispersion $\sglos$,
considering particles in a slit of width $\Re/4$ around the semi-major
axis of the isophotal ellipse. Figure~\ref{fig2d} plots $\vlos$ (solid
line) and $\sglos$ (dashed line) of the stellar component for four
systems with different combinations of dark matter mass $\Mh$ and spin
parameter $\lambda$. The projected velocity dispersion profiles do not
present the central depressions characterizing the isotropic Sersic
profiles or the Hernquist model. In the presence of significant
angular momentum (bottom panels) $\vlos$ becomes comparable with
$\sglos$ at $R \sim {\rm few}\,\Re$ when the line-of-sight direction
and the angular momentum vector are orthogonal.

\begin{figure}
\psfig{file=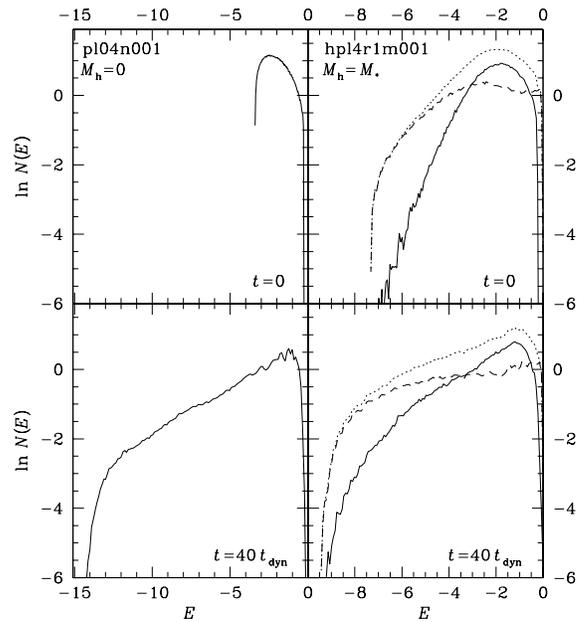,width=\hsize}
 \caption{Left panels: initial (top) and final (bottom) differential
 energy distribution for a representative one-component simulation.
 Right panels: same as left panels for a two-component
 simulation. Solid, dashed and dotted curves refer to the stellar,
 halo, and total distributions, respectively. The energy per unit mass
 $E$ is normalized to $\Etot/\Mtot$.}
\label{fignde}
\end{figure}

\subsection{Differential energy distribution}

In Fig.~\ref{fignde} (left panels) we plot the initial (top) and final
(bottom) differential energy distributions $N(E)$ (i.e. the number of
particles with energy per unit mass between $E$ and $E+dE$) for a
representative one-component collapse simulation. The final
differential energy distribution is well represented by an exponential
$N(E)\propto \exp{(\eta E)}$, over most of the energy range, with
$\eta\sim 2.1$ (when the energy per unit mass is normalized to $G
\Mstar / \Re$, see Binney 1982; van Albada 1982; Ciotti~1991).  We
find that the colder the initial conditions, the larger the range in
$E$ in which $N(E)$ is exponential, because only very cold initial
conditions produce strongly bound particles (and, correspondingly,
populate the very central regions; see Section~\ref{secbeh}).

In the right panels we plot, for a representative two-component
collapse, the initial and final $N(E)$ of the stellar (solid lines)
and halo (dashed lines) component, together with the total $N(E)$
(dotted lines).  Note how the energy distribution of the final stellar
component is steeper than the halo component, i.e., how the very-bound
end of the energy distribution (typically corresponding to particles
in the central region of the density distribution) is populated by
halo particles (dashed line), while most of the stellar particles are
close to $E \sim 0$.  Interestingly, the slope of the total
differential energy distribution (dotted line) is very similar to that
found in the one-component case when $E$ is normalized to
$\Etot/\Mtot$.

\subsection{Exploring different initial conditions}

In general, the large-scale properties of the end-products of
one-component simulations do not depend strongly on the initial
distribution, provided that it is clumpy and cold enough.  However,
some information on the initial distribution is not completely erased
during the collapse. As apparent from Fig.~\ref{figmem}, while Plummer
(blue symbols) initial conditions lead to systems very close to the
de Vaucouleurs law ($3.6\lsim m \lsim 5.2$), the end-products of
$\gamma=0$ (green symbols) and $\gamma=1$ (red symbols) initial
distributions are characterized by higher Sersic indices ($5\lsim m
\lsim 8$).

In two-component simulations the initial shape of the dark matter halo
does not have strong influence on the properties of the stellar
end-product.  In particular, for a wide range of halo density profiles
($\gamma=0$, $\gamma=1$, $\gamma=1.5$, and NFW distributions), the
final stellar density profiles have $m\lsim\,4$, down to 2, if the
halo is sufficiently concentrated.  If the halo is light and/or
extended, independently of its shape, the final stellar components
have $m\gsim\,4$. We also considered cases in which the halo, as well
as the stellar component, is strongly out of equilibrium ($\betah \ll
1$; see Table~1).  Independently of the details of their initial halo
density distributions, the stellar end-products of these simulations
have global properties more similar to one-component collapses than to
collapses in nearly virialized haloes. As a consequence, these
end-products (represented as stars in Fig.~\ref{figmem}), contribute
significantly to the scatter in the left diagram of Fig.~\ref{figmem},
because they typically have higher values of $m$ than the collapses in
nearly virialized haloes with similar dark-to-luminous mass ratio.
In addition, in analogy with the one-component case,
two-component collapses with $\gamma=1$ initial stellar distribution
(such as g0hq05r2m002, red filled circles in Fig.~\ref{figmem})
produce systems with higher Sersic index than those with Plummer
initial distribution (blue symbols) with similar dark-to-luminous mass
ratio (see Fig.~\ref{figmem}, left panel). Interestingly, the stellar
system with the highest Sersic index ($m\simeq 12.1$) was obtained
starting with a $\gamma=1$ stellar distribution embedded in a
collapsing dark matter halo (simulation g0hq05r1m005, red stars in
Fig.~\ref{figmem}).

We also considered a rather extreme case of two-component initial
conditions (model hus20r1m002 in Table~1, magenta curves and symbols in
Fig.~\ref{figden},~\ref{figmem}, and \ref{figaxes}) that could be also
interpreted as a multiple merging experiment. The halo is a Hernquist
model with core radius $\rh$, and the stars are distributed in 20
uniform spheres of radius $2\rh$, whose centres of mass are uniformly
distributed in a sphere of radius $20\rh$. Each small sphere is cold,
having all particles at rest with respect to its centre of mass. So
the small spheres collapse first because of self-gravity and then
merge because of dynamical friction, finally forming a stellar system
with best-fit Sersic index\footnote{Similar values of $m$ were also
found by Nipoti et al. (2003b) for the end-products of galactic
cannibalism simulations.}  $m\sim6.5$. We note that in this case the
final central halo density distribution is significantly flatter than
the the initial one, as expected as a consequence of dynamical
friction heating (see also Nipoti et al.~2004).

According to these results, it appears that the only way to produce
$m\gsim 5$ systems with dissipationless processes is either merging or
collapse starting from concentrated and cold initial distributions,
while dissipationless collapses of more diffuse, cold distributions
lead to $m\lsim 5$ systems. We note that lower values of $m$ were also
found as a result of quite different dissipationless processes:
Merritt et al.~(2005) found that the projected density of the dark
matter haloes of cosmological N-body simulations are well fitted by
the Sersic law with $2\lsim m \lsim 4$, while Nipoti et al.~(2002)
obtained $1\lsim m \lsim 5$ for the end-products of radial orbit
instability.

\begin{figure}
\psfig{file=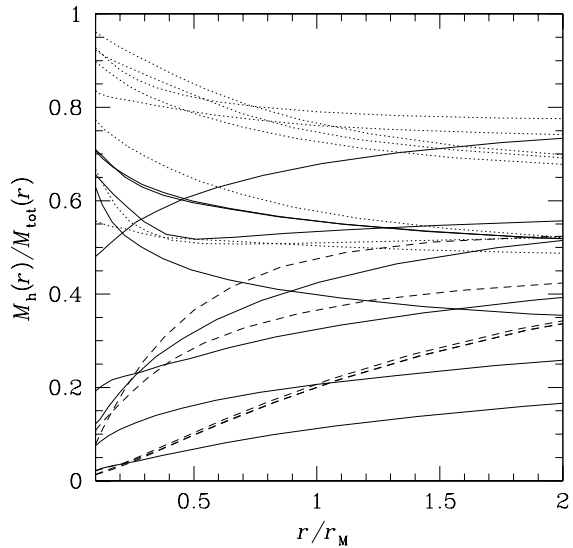,width=.9\hsize}
 \caption{Dark-to-total mass inside $r$ for the end-products of the
 two-component simulations. Dotted, solid, and dashed curves refer to
 systems with average best-fitting Sersic index $\mav \le 3$, $3 <\mav
 < 5$, and $\mav \ge 5$, respectively [$\mav=(m_a+m_b+m_c)/3$].}
\label{figmrat}
\end{figure}

\section{Discussion and conclusions}

Before summarizing the main results of the numerical simulations, it
is useful to list three (of the several) {\it observational}
properties of ellipticals that should be explained by any plausible
formation (or evolutionary) scenario:

\begin{enumerate}

\item The best-fit Sersic index increases from $m \sim 2$ to $m \sim
10$ from faint to luminous galaxies (Caon et al.~1993; Bertin et
al.~2002). Thus, the light profiles of luminous ellipticals as a
function of $R/\Re$ are {\it steeper} for $R \lsim\,\Re$ and flatter
for $R \gsim\,\Re$ than the profiles of low-luminosity ellipticals
(see, e.g., Ciotti~1991).

\item However, the light profiles of luminous ellipticals present a
flat core in their very central regions, deviating from the global
best-fit Sersic law. The profiles of low-luminosity ellipticals are
instead well fitted by the Sersic law down to the centre (e.g. Lauer
et al. 1995; Trujillo et al. 2004), so in the very central regions
luminous ellipticals have flatter density distributions than
low-luminosity ellipticals .

\item Finally, luminous ellipticals have lower mean density within
$\Re$ than low-luminosity ellipticals, a well known consequence of the
Kormendy~(1977) relation.

\end{enumerate}

In this paper we presented the results of numerical N-body simulations
of one- and two-component dissipationless collapses, focusing on their
possible relevance to the origin of the weak homology and of the
core/power-law dichotomy of ellipticals, i.e., we addressed points (i)
and (ii) above, while it is quite obvious that point (iii) cannot be
studied with dissipationless collapse simulations.  Overall, our
results suggest that dissipationless collapse is effective in
producing weak structural homology, kinematic properties consistent
with observations, and cores on scales remarkably similar to those
observed. In particular:

\begin{itemize}

\item The end-products of one-component dissipationless collapses
  typically have projected surface brightness profile close to the de
  Vaucouleurs model (see also vA82; Londrillo et al.~1991; Dantas et
  al.~2002; Trenti et al.~2004). When fitted with the Sersic law over
  the radial range $0.1 \, \lsim \, R/\Re \, \lsim \, 10$, the
  resulting profiles are characterized by index $3.6\lsim m \lsim
  8$; $m \gsim 5$ final states are obtained only for rather
  concentrated initial conditions.

\item The end-products of collapses inside a dark matter halo present
  significant structural non-homology. The best-fit Sersic indices of
  the stellar projected surface density profile span the range
  $1.9\lsim m \lsim 12$. Remarkably, the parameter $m$ correlates
  with the amount of dark matter present within $\Re$, being smaller
  for larger dark-to-visible mass ratios.

\item The projected stellar density profiles are characterized by a
  break radius $0.01\lsim \Rb/\Re \lsim0.1$ within which the profile
  is flatter than the inner extrapolation of the global best-fit
  Sersic law. Colder initial conditions lead to end-products with
  smaller $\Rb/\Re$; in general, the resulting `cores' are better
  detectable in high-$m$ systems.

\end{itemize}

The results above indicate that dissipationless collapse is able to
produce systems with projected density profiles remarkably similar to
the observed surface brightness profiles, with high quality
one-parameter Sersic fits even for low $m$ values when non-negligible
amounts of dark matter are present. In particular, in
Fig.~\ref{figmrat} we plot the (integrated) dark-to-total mass ratio
$\Mh(r)/\Mtot(r)$ for the two-component simulations end-products. A
clear dichotomy is apparent: while in low-$m$ systems (dotted lines)
the mass-to-light ratio increases significantly inwards, the opposite
is true for systems with $m\gsim3$ (with few exceptions). We also note
that, while for $m\gsim3$ systems the dark matter mass inside the half
mass radius is of the same order as the visible mass, in accordance
with observations (e.g. Bertin et al.~1994; Gerhard et al.~2001;
Magorrian \& Ballantyne~2001; Romanowsky et al.~2003; Treu \&
Koopmans~2004; Cappellari et al.~2006; Dekel et al.~2005; Samurovi\'c
\& Danziger 2005), for very low-$m$ systems the dark matter is
dominant.

According to the standard interpretation, the central cores observed
in several bright ellipticals are a consequence of formation through
merging, being produced by the interaction of binary supermassive
black holes with a stellar cusp (Makino \& Ebisuzaki 1996; Faber et
al. 1997; Milosavljevic \& Merritt 2001).  Our results, in agreement
with previous numerical explorations (vA82; May \& van Albada 1984),
indicate that a break in the profile at small radii and a flat central
core are features produced naturally by dissipationless collapse.  In
particular, the fact that the size of the core is correlated (and the
maximum central density anti-correlated) with the initial virial ratio
is a direct consequence of Liouville Theorem\footnote{Hozumi,
Burkert \& Fujiwara (2000) pointed out that another consequence of the
conservation of the phase-space density is that the initial anisotropy
in the velocity distribution affects the final density profile at
small radii. As shown by their spherically-symmetric dissipationless
collapse simulations, smaller central cores are produced by more
radially anisotropic collapses.} (e.g., May \& van Albada 1984). The
higher resolution of our simulations allowed us to investigate this
effect in regions (down to $R\sim0.01\Re$) comparable to those
explored by high-resolution photometry of real ellipticals.  In
particular, the profiles of the end-products of our single-component
simulations are remarkably similar to those of observed `core'
ellipticals. For instance, it is suggestive to compare the profile
plotted in Fig.~\ref{figinn} (bottom, left panel) with that of the
core-elliptical NGC\,3348 (see fig.~10 of Graham et al.~2003), for
which Trujillo et al.~(2004) report best-fit $m\simeq 3.8$ and
$\Rb/\Re\simeq0.016$.  The fact that the end-products of our
simulations reproduce very nicely the observed cores makes
dissipationless collapse a plausible alternative to the binary black
hole scenario for the origin of the cores. The next step of the
present study would be the exploration with a hydro N-body code of the
first stages of galaxy formation dominated by gaseous dissipation and
the following feedback effects due to star formation.


\section*{Acknowledgments}

We thank the anonymous Referee for useful comments.  L.C and P.L. were
supported by the MIUR grant CoFin~2004.  C.N. and P.L. are grateful to
CINECA (Bologna) for assistance with the use of the IBM Linux Cluster.


\begin{thebibliography}{99}

\bibitem{} Bender R., Burstein D., Faber S. M., 1993, ApJ, 411, 153
\bibitem{}Bertin G., Ciotti L., Del Principe M., 2002, A\&A, 386, 1491
\bibitem{}Bertin G., et al. 1994, A\&A, 292, 381
\bibitem{}Binney J., 1982, MNRAS, 200, 951
\bibitem{}Burstein D., Davies R.L., Dressler A., Faber S.M.,
Lynden-Bell D., 1988. In: {\it Towards understanding galaxies at large
redshift; Proceedings of the Fifth Workshop of the Advanced School of
Astronomy}, Dordrecht, Kluwer Academic Publishers.
\bibitem{}Cappellari M., et al., 2006, MNRAS, 366, 1126
\bibitem{}Ciotti L., 1991, A\&A, 249, 99
\bibitem{}Ciotti L., 1996, ApJ, 471, 68
\bibitem{}Ciotti L., 1999, ApJ, 520, 574
\bibitem{}Ciotti L., Bertin G., 1999, A\&A, 352, 447
\bibitem{}Ciotti L., Pellegrini S., 1992, MNRAS, 255, 561
\bibitem{}Ciotti L., van Albada, T.S., 2001, ApJ, 552, L13 
\bibitem{}Caon N., Capaccioli M., D'Onofrio M., 1993, 265, 1013 
\bibitem{}Dantas C.C., Capelato H.V., de Carvalho R.R., Ribeiro A.L.B., 2002, A\&A, 384, 772
\bibitem{}Dehnen W., 1993, MNRAS, 265, 250
\bibitem{}Dehnen W., 2000, ApJ, 536, L39
\bibitem{}Dehnen W., 2002, Journal of Computational Physics, 179, 27
\bibitem{}Dekel A., Stoehr F., Mamon G.A., Cox T.J., Novak G.S., Primack J.R., 2005, Nature, 437, 707
\bibitem{}de Vaucouleurs G., 1948, Ann. d'Astroph., 11, 247
\bibitem{} Eggen O.J., Lynden-Bell D., Sandage A.R., 1962, ApJ, 136,
748
\bibitem{}Faber S.M., Jackson R.E., 1976, ApJ, 204, 668
\bibitem{}Faber S.M., et al., 1997, AJ, 114, 1771
\bibitem{}Ferrarese L., Merritt D., 2000, ApJ, 539, L9
\bibitem{}Ferrarese L., van den Bosch F.C., Ford H.C., Jaffe W.,
  O'Connell R.W., 1994, AJ, 108, 1598
\bibitem{}Ferrarese L., et al., 2006, ApJS, in press (astro-ph/0602297)
\bibitem{} Gebhardt K. et al., 2000, ApJ, 539, L13
\bibitem{} Gerhard O., Kronawitter A., Saglia R.P., Bender R., 2001, 
           AJ, 121, 1936
\bibitem{}Graham A.W., Guzm\'an R., 2003, AJ, 125, 2936
\bibitem{}Graham A.W., Erwin P., Trujillo I., Asensio Ramos A., 2003, ApJ,
125, 2951
\bibitem{}Hernquist L., 1990, ApJ, 356, 359 
\bibitem{}Hozumi S., Burkert A., Fujiwara T.,  2000, MNRAS, 311, 377
\bibitem{}Kormendy J., 1977, ApJ, 218, 333
\bibitem{}Lake G., 1983, ApJ, 264, 408
\bibitem{}Lauer T.R., et al., 1995, AJ, 110, 2622
\bibitem{}Londrillo P., Messina A., Stiavelli M., 1991, MNRAS, 250, 54
\bibitem{}Londrillo P., Nipoti C., Ciotti L., 2003, In
``Computational astrophysics in Italy: methods and tools'', Roberto
Capuzzo-Dolcetta ed., Mem. S.A.It. Supplement, vol. 1, p. 18
\bibitem{}Magorrian J.,  Ballantyne, D., 2001, MNRAS, 322, 702
\bibitem{}May A.,  van Albada, T.S., 1984, MNRAS, 209, 15
\bibitem{}Makino J., Ebisuzaki T., 1996, ApJ, 465, 527
\bibitem{}McGlynn T.A., 1984, ApJ, 281, 13
\bibitem{}Merritt D., Navarro J.F., Ludlow A., Jenkins A.,  2005,  ApJ, 624, L85\bibitem{}Milosavljevic M., Merritt D., 2001, ApJ, 563, 34
\bibitem{}Naab T., Johansson P.H., Efstathiou G., Ostriker J.P., 2005,
  submitted to ApJ (astro-ph/0512235)
\bibitem{}Navarro J.F., Frenk C.S., White S.D.M., 1996, ApJ, 462, 563 (NFW)
\bibitem{}Nipoti C., Londrillo P., Ciotti L., 2002, MNRAS, 332, 901 
\bibitem{}Nipoti C., Londrillo P., Ciotti L., 2003a, MNRAS, 342, 501 (NLC03)
\bibitem{}Nipoti C., Stiavelli M., Treu T., Ciotti L., Rosati P., 2003b, MNRAS, 344, 748
\bibitem{}Nipoti C., Treu T., Ciotti L., Stiavelli M.,  2004, MNRAS, 355, 1119
\bibitem{}Pellegrini S., 1999, A\&A, 351, 487
\bibitem{}Pellegrini S., 2005, MNRAS, 364, 169
\bibitem{}Power C., Navarro J.F., Jenkins A., Frenk C.S., White S.D.M., Springel V., Stadel J., Quinn T., 2003, MNRAS, 338, 14
\bibitem{}Plummer H.C., 1911, MNRAS, 71, 460 
\bibitem{}Prugniel P., Simien F., 1997, A\&A, 321, 111
\bibitem{}Robertson B., Cox T.J., Hernquist L., Franx M., Hopkins
  P.F., Martini P., Springel V., 2006, ApJ, 641, 21
\bibitem{}Romanowsky A.J., Douglas N.G., Arnaboldi M., Kuijken K.,
  Merrifield M.R., Napolitano N.R., Capaccioli M., Freeman K.C., 2003,
  Science, 301, 1696
\bibitem{}Samurovi\'c S., Danziger I.J., 2005, MNRAS, 363, 769
\bibitem{}Sersic J.L., 1968, Atlas de galaxias australes. Observatorio Astronomico, Cordoba
\bibitem{}Tremaine S., Richstone D.O., Yong-Ik B., Dressler A., Faber S.M., 
           Grillmair C., Kormendy J., Laurer T.R., 1994, AJ, 107, 634
\bibitem{}Trenti M., Bertin G., van Albada T.S., 2005, A\&A, 433, 57
\bibitem{}Treu T., Koopmans L.V., 2004, ApJ, 611, 739
\bibitem{}Trujillo I., Erwin P., Asensio Ramos A., Graham A.W., 2004, AJ, 127, 1917
\bibitem{}Udry S., 1993, A\&A, 268, 35
\bibitem{}van Albada T.S., 1982, MNRAS, 201, 939 (vA82)
\bibitem{}White~S.D.M., Frenk~C.S., 1991, ApJ, 379, 52
\bibitem{}White~S.D.M., Rees~M.J., 1978, MNRAS, 183, 341
\end{thebibliography}
\end{document}